\documentstyle[fullpage,11pt,epsf]{article}

\title{Transfer-Matrix Study of Hard-Core Singularity \\ 
for Hard-Square Lattice Gas}

\author{Synge Todo \\
{\it Institute for Solid State Physics, University of Tokyo,} \\
{\it Roppongi, Minato-ku, Tokyo 106, Japan}}

\begin{document}

\maketitle

\begin{abstract}
A singularity on the negative fugacity axis of the hard-square
  lattice gas is investigated in terms of numerical diagonalization of
  transfer matrices.  The location of the singular point $z_{\rm c}^-$
  and the critical exponent $\nu$ are accurately determined by the
  phenomenological renormalization technique as $-0.11933888188(1)$
  and $0.416667(1)$, respectively.  It is also found that the central
  charge $c$ and the dominant scaling dimension $x_\sigma$ are
  $-4.399996(8)$ and $-0.3999996(7)$, respectively.  These results
  strongly support that this singularity belongs to the same
  universality class as the Yang-Lee edge singularity
  ($c=-\frac{22}{5}$, $x_\sigma=-\frac{2}{5}$ and
  $\nu=\frac{5}{12}$).
\end{abstract}

\section{Introduction}

For systems of hard particles, the radius of convergence of the
fugacity series expansion is known to be generally determined by a
{\em non-physical} singularity on the negative real fugacity $z$
axis~\cite{Groeneveld62,GauntF65,Gaunt67,Gaunt69}. Here, let us
consider the hard-square lattice gas (hard squares), which is one of
the simplest models describing the fluid-solid transition in two
dimensions.  This model exhibits a second-order transition at
\begin{equation}
  z_{\rm c}^{+}=3.796255174(3)
\end{equation}
from the disordered phase to the $\sqrt{2}\times\sqrt{2}$
phase~\cite{BloteW90} (uncertainty in the last decimal digits is given
by the figure in parentheses).  It is also precisely confirmed that
the transition belongs to the same universality class as the Ising
ferromagnet~\cite{TodoS96}.

On the other hand, the alternating behavior in the sign of the
coefficients in the fugacity series of the reduced
pressure~\cite{BaxterET80}:
\begin{eqnarray}
  \label{eqn:series}
  P &=& z - \frac{5}{2} z^2 + \frac{31}{3} z^3 - \frac{209}{4} z^4 
  + \frac{1476}{5} z^5 - \frac{10739}{6} z^6 \nonumber \\
  & & \ \mbox{} + \frac{79780}{7} z^7 - \frac{601905}{8} z^8
  + \frac{4595485}{9} z^9
  - \cdots \nonumber \\
  & & \mbox{}
\end{eqnarray}
manifests that a {\em dominant} singularity exists on the negative
fugacity axis.  By series analysis~\cite{Guttmann87}, the location of
the singularity is estimated as
\begin{equation}
  \label{eqn:guttmann}
  z_{\rm c}^{-} = -0.119 338 880 9 (10) .
\end{equation}
Note that the absolute magnitude of $z_{\rm c}^{-}$ is about 30 times
smaller than that of $z_{\rm c}^{+}$.  This fact makes it quite
difficult to derive precise information around the {\em physical}
critical point $z_{\rm c}^{+}$ from the fugacity
series~\cite{GauntF65}. Similar situations are commonly observed in
other hard-core systems.

This non-physical singularity is referred to as the {\em hard-core
singularity} or the {\em repulsive-core singularity}.  At this point
thermodynamic quantities are known to exhibit interesting non-trivial
behavior.  For example, the reduced pressure or equivalently the
reduced free-energy density behaves as
\begin{equation}
  -P(z) = f(z) \sim (z-z_{\rm c}^{-})^\phi
\end{equation}
for $z \rightarrow z_{\rm c}^{-}+$.  The singularity is characterized
by the exponent $\phi$.  For the hard-square lattice gas $\phi$ is
found to be about $0.83337$ by series analysis~\cite{Guttmann87}.

In 1984, Poland~\cite{Poland84} investigated the fugacity series for a
variety of hard-core lattice gases and hard-core gases in continuous
space in two, three and higher dimensions, and proposed that the
exponent $\phi$ is {\em universal}, that is, it depends only on the
dimensionality of space.  Another study on further models by Baram and
Luban~\cite{BaramL87} also supports this conjecture.

In addition, Lai and Fisher~\cite{LaiF95} pointed out that the
hard-core singularity can be identified with the {\em Yang-Lee edge
  singularity}~\cite{YangL52,LeeY52,KortmanG71,Fisher78,Cardy85,ItzyksonSZ86}
(the exponent $\phi$ of the hard-core singularity relates to Fisher's
exponent $\sigma$ of the Yang-Lee edge singularity by
$\phi=\sigma+1$).  It means, if it is true, that the singular point
$z=z_{\rm c}^-$ can be regarded as a conventional critical point; the
correlation length diverges and consequently ordinary finite-size
scaling analysis does work.

As far as we know, all the studies with respect to the hard-core
singularity are done by means of series analysis up to the present
time (except for the analytic solution for hard
hexagons~\cite{BaramL87,Baxter80} and the trivial cases in dimensions
less than two).  In this paper, we investigate this singularity of the
hard-square lattice gas by another approach --- numerical
diagonalization of transfer matrices and phenomenological
renormalization analysis.

The transfer-matrix method is one of the most powerful numerical
methods in investigating statistical mechanical models in low
dimensions~\cite{Nightingale90}. It has some advantages such as the
absence of statistical errors and critical slowing down, which are
observed in Monte Carlo simulations.  Moreover, in two dimensions, the
conformal invariance associated with a critical system yields a great
deal of interesting
consequences~\cite{BelavinPZ84,FriedanQS84,Cardy87}. Especially, useful
asymptotically exact relations between universal quantities, such as
the critical exponent, and eigenvalues of the transfer matrix are
given.  This enables us to perform precise analysis about the
criticality even in the non-unitary case with a negative central
charge, such as the present hard-core singularity, as shown later.

The organization of the present paper is as follows: In \S 2 we
describe the transfer-matrix method and give the explicit notation of
the transfer matrix for the hard-square lattice gas.  In \S 3 the
critical point of the hard-core singularity and some universal
quantities are investigated by means of the phenomenological
renormalization technique.  In addition, in \S 4 we consider the
critical eigenvalue spectrum of the transfer matrix precisely, which
relates to the operator content of the corresponding conformal theory.
We give a summary and some discussion in the final section.

%%%%%%%%%%%%%%%%%%%%%%%%%%%%%%%%%%%%%%%%%%%%%%%%%%%%%%%%%%%%%%%%%%%%
\section{Transfer Matrix for Hard-Square Lattice Gas}

The partition function of the hard-square lattice gas is defined as
follows:
\begin{equation}
  \label{eqn:partition}
  Z = \sum_{\{s_i\}} z^{\sum_i s_i}
  \prod_{\langle i,j \rangle} (1-s_i s_j) ,
\end{equation}
where $s_i$ is a one-bit binary number, which describes whether the
$i$th site on the square lattice is occupied ($s_i=1$) or vacant
($s_i=0$).  The product in eq.~(\ref{eqn:partition}) runs over all
nearest-neighbor pairs of the lattice sites.  The size of hard squares
is so that any two of them can not occupy on nearest-neighbor
positions simultaneously.

The partition function of the system of $L \times M$ sites with
periodic boundary conditions can be expressed in terms of the transfer
matrix~\cite{Nightingale90}:
\begin{equation}
  Z_{LM} = {\rm Tr} \ (T_L)^M .
\end{equation}
The transfer matrix $T_L$ is square and of order $2^L$.  Note there
remains some options in choosing the {\em transfer direction} and
dividing the total Hamiltonian into a sum of {\em slice
Hamiltonians}~\cite{Nightingale90}. In this paper, we consider the
following two types of transfer matrices, which have distinct transfer
directions with each other.

%%%%%%%%%%%%%%%%%%%%%%%%%%%%%%%%%%%%%%%%%%%%%%%%%%%%%%%%%%%%%%%%%%%%
\subsection{Row-to-row transfer matrix}

\begin{figure}[t]
\begin{center}
  \leavevmode
  \epsfysize=2.0truein \epsffile{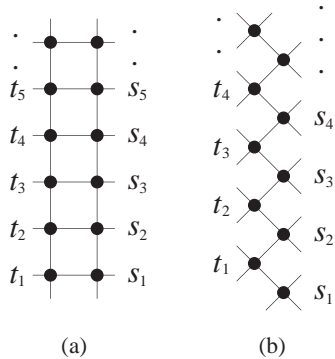}
\caption{Two neighboring unit slices of the row-to-row~(a) 
  and the diagonal-to-diagonal~(b) transfer matrices.  The transfer
  direction is chosen in the horizontal direction (right to left) in
  both cases.  In the vertical direction, periodic boundary conditions
  are applied.}
\label{fig:lattice}
\end{center}
\end{figure}

The first one is the {\em row-to-row} transfer matrix.  The transfer
direction is chosen to be parallel to a set of lattice edges
(Fig.~\ref{fig:lattice}~(a)).  The $(\tau|\sigma)$ element is
explicitly written as
\begin{equation}
  \label{eqn:tm_rr}
  T_{{\rm r},L} (\tau | \sigma) =
  z^{\sum_{i=1}^L s_i}
  \prod_{i=1}^L (1-s_i s_{i+1}) (1-s_i t_i) (1-t_i t_{i+1}) ,
\end{equation}
where $\sigma=(s_1,s_2,\cdots,s_L)$ and $\tau=(t_1,t_2,\cdots,t_L)$
respectively denote one of $2^L$ configurations on the two neighboring
unit slices respectively ($s_i$, $t_i=1$ or 0).  Periodic boundary
conditions along the unit slices (the vertical direction in
Fig.~\ref{fig:lattice}~(a)) are applied ($s_{i+L}=s_i$ and
$t_{i+L}=t_i$).

It should be noticed that quite a number of matrix elements should
vanish owing to the presence of the hard-core interaction.
Especially, the number of rows or columns containing at least one
non-zero element is only $\left(\frac{1}{2}(1+\sqrt{5})\right)^L =
(1.61803\cdots)^L$ for $L \gg 1$.  This is much smaller than $2^L$.
One can construct an algorithm for multiplication of the matrix to a
trial vector almost in this highly restricted space as shown by Todo
and Suzuki~\cite{TodoS96}. Thus, dominant eigenvalues of the transfer
matrix can be calculated effectively up to the system width $L=28$,
which is quite larger than that achieved by the standard sparse-matrix
factorization~\cite{BloteW90,BloteWW90}.

%%%%%%%%%%%%%%%%%%%%%%%%%%%%%%%%%%%%%%%%%%%%%%%%%%%%%%%%%%%%%%%%%%%%
\subsection{Diagonal-to-diagonal transfer matrix}

We also consider the {\em diagonal-to-diagonal} transfer matrix.  In
this case, the transfer direction is rotated by the amount of
$\frac{\pi}{4}$ to a set of lattice edges
(Fig.~\ref{fig:lattice}~(b)).  Periodic boundary conditions are also
applied in the orthogonal direction.  Matrix elements are written as
follows:
\begin{equation}
  \label{eqn:tm_dd}
  T_{{\rm d},L} (\tau | \sigma) =
  z^{\sum_{i=1}^L s_i}
  \prod_{i=1}^L (1-s_i t_i) (1-s_{i+1} t_i) ,
\end{equation}
In this case, any two particles on the sites in a unit slice do not
interact with each other.  Consequently, all the rows or columns of
the matrix $T_{{\rm d},L}$ contains at least one none-zero element.
By using the standard sparse-matrix factorization
technique~\cite{BloteW90,BloteWW90}, numerical matrix multiplication
to a vector can be carried out up to $L=20$.

%%%%%%%%%%%%%%%%%%%%%%%%%%%%%%%%%%%%%%%%%%%%%%%%%%%%%%%%%%%%%%%%%%%%
\subsection{Eigenvalues, free-energy density and correlation length}

In the $M \rightarrow \infty$ limit, the reduced free-energy density
$f_{L}$ is given in terms of the largest eigenvalue $\lambda_1$ of the
transfer matrix as
\begin{eqnarray}
  \label{eqn:free}
  f_L &=& \lim_{M\rightarrow\infty} 
  \left(-\frac{1}{LM} \log Z_{LM} \right) \nonumber \\
  &=& -\frac{1}{L} \log \lambda_1 .
\end{eqnarray}
The inverse correlation length in the transfer direction, which is
also referred to as the {\em gap}, is calculated as
\begin{equation}
  \label{eqn:gap}
  \xi_L^{-1} = g_L = \log \left| \frac{\lambda_1}{\lambda_2} \right| ,
\end{equation}
where $\lambda_2$ is the second largest eigenvalue in absolute magnitude.  

It should be noted that for $z<0$ a number of matrix elements in
eqs.~(\ref{eqn:tm_rr}) and (\ref{eqn:tm_dd}) become negative.
Consequently, there is no guarantee that the largest eigenvalue is
unique.  In general, it may appear as a member of degenerating
eigenvalues or one of a complex-conjugate pair.  It is in a sharp
contrast to the case for $z>0$, where the largest eigenvalue is always
real and non-degenerate as a consequence of the Perron-Frobenius
theorem~\cite{Gantmacher59}.

However, in the present case, one can easily show that there exists a
certain fugacity region ($z_L^\times <z<0$), where the largest
eigenvalue is actually real and non-degenerate, as the following:

First of all, note that all the elements of $T_{{\rm r},L}$ and
$T_{{\rm d},L}$ are real. Then, it is apparent coefficients of their
characteristic polynomials are all real.  This means that imaginary
eigenvalues should appear as complex-conjugate pairs, if they exist.
On the other hand, it can be easily seen that at $z=0$ the largest
eigenvalue is unity and all the other eigenvalues are zero.
Consequently, the largest eigenvalues is real and non-degenerate at
least between the $z=0$ and the point $z=z_L^\times$, where the
largest and the second-largest eigenvalues first cross with each
other~\cite{fn1}. For $z<z_L^\times$, they form a complex-conjugate
pair.  In the following, we work only in the region $z_L^\times < z <
0$.

\begin{table}[t]
\caption{Critical point $z_L^*$, exponent $\nu_L$, scaled gap 
  $Lg_L(z)/2\pi$ at $z_L^*$ and central charge $c_L$ for the
  row-to-row transfer matrix.  Their extrapolated values to
  $L\rightarrow\infty$ are also listed in the bottom row.}
\label{tab:row}
\begin{center}
\begin{tabular}{rllll}
  \hline
  $L$ & 
  \multicolumn{1}{c}{$z^*_L$} & 
  \multicolumn{1}{c}{$\nu_L$} & 
  \multicolumn{1}{c}{$L g_L/2\pi$} & 
  \multicolumn{1}{c}{$c_L$} \\
  \hline
  2  & -0.11569699529389 & 0.421849529 & 0.4814837782 & 0.417414138 \\
  3  & -0.11856124181461 & 0.416582977 & 0.4394700591 & 0.438343788 \\
  4  & -0.11908596327934 & 0.415750210 & 0.4233109453 & 0.430386102 \\
  5  & -0.11923864086390 & 0.415659639 & 0.4150498994 & 0.421997003 \\
  6  & -0.11929304185438 & 0.415705242 & 0.4104150242 & 0.416258939 \\
  7  & -0.11931538407218 & 0.415775276 & 0.4076315079 & 0.412555119 \\
  8  & -0.11932569455336 & 0.415846360 & 0.4058502112 & 0.410099405 \\
  9  & -0.11933093621984 & 0.415912883 & 0.4046435075 & 0.408387975 \\
  10 & -0.11933381954516 & 0.415973444 & 0.4037860694 & 0.407135062 \\
  11 & -0.11933550958958 & 0.416027918 & 0.4031528434 & 0.406179637 \\
  12 & -0.11933655239965 & 0.416076654 & 0.4026705275 & 0.405427508 \\
  13 & -0.11933722350045 & 0.416120179 & 0.4022938659 & 0.404820630 \\
  14 & -0.11933767079976 & 0.416159066 & 0.4019936087 & 0.404321316 \\
  15 & -0.11933797791846 & 0.416193868 & 0.4017501012 & 0.403903960 \\
  16 & -0.11933819423819 & 0.416225088 & 0.4015496991 & 0.403550501 \\
  17 & -0.11933835002255 & 0.416253176 & 0.4013826680 & 0.403247800 \\
  18 & -0.11933846442079 & 0.416278521 & 0.4012418994 & 0.402986066 \\
  19 & -0.11933854989180 & 0.416301464 & 0.4011220978 & 0.402757844 \\
  20 & -0.11933861474296 & 0.416322296 & 0.4010192489 & 0.402557360 \\
  21 & -0.11933866463558 & 0.416341268 & 0.4009302622 & 0.402380074 \\
  22 & -0.11933870350381 & 0.416358598 & 0.4008527257 & 0.402222367 \\
  23 & -0.11933873413008 & 0.416374471 & 0.4007847341 & 0.402081322 \\
  24 & -0.11933875851387 & 0.416389050 & 0.4007247649 & 0.401954561 \\
  25 & -0.11933877811309 & 0.416402474 & 0.4006715898 & 0.401840128 \\
  26 & -0.11933879400494 & 0.416414865 & 0.4006242084 & 0.401736403 \\
  27 & -0.11933880699512 & 0.416426329 & 0.4005817991 & 0.401642027 \\
  \hline
  $\infty$ & -0.11933888189(2) & 0.416668(2) & 0.3999996(7) & 0.400000(2) \\
  \hline
\end{tabular}
\end{center}
\end{table}

\begin{table}[t]
\caption{Critical point $z_L^*$, exponent $\nu_L$, scaled gap 
  $Lg_L(z)/2\pi$ at $z_L^*$ and central charge $c_L$ for the
  diagonal-to-diagonal transfer matrix.  Their extrapolated values
  to $L\rightarrow\infty$ are also listed in the bottom row.}
\label{tab:diagonal}
\begin{center}
\begin{tabular}{rllll}
  \hline
  $L$ & 
  \multicolumn{1}{c}{$z^*_L$} & 
  \multicolumn{1}{c}{$\nu_L$} & 
  \multicolumn{1}{c}{$L g_L/2\pi$} & 
  \multicolumn{1}{c}{$c_L$} \\
  \hline
  2  & -0.11876248705617 & 0.447460528 & 0.4308877716 & 0.320446273 \\
  3  & -0.11923273442241 & 0.431550594 & 0.4126164837 & 0.355461268 \\
  4  & -0.11930719600997 & 0.425758018 & 0.4068301638 & 0.370796613 \\
  5  & -0.11932661625882 & 0.422898114 & 0.4042554292 & 0.379106387 \\
  6  & -0.11933327955420 & 0.421248828 & 0.4028891193 & 0.384181172 \\
  7  & -0.11933601031532 & 0.420200952 & 0.4020795070 & 0.387534567 \\
  8  & -0.11933727916302 & 0.419489095 & 0.4015617111 & 0.389879246 \\
  9  & -0.11933792672807 & 0.418981068 & 0.4012113921 & 0.391590106 \\
  10 & -0.11933828221945 & 0.418604487 & 0.4009639003 & 0.392880960 \\
  11 & -0.11933848910493 & 0.418316797 & 0.4007829378 & 0.393881552 \\
  12 & -0.11933861540512 & 0.418091545 & 0.4006468626 & 0.394674591 \\
  13 & -0.11933869564988 & 0.417911541 & 0.4005421319 & 0.395314964 \\
  14 & -0.11933874838811 & 0.417765191 & 0.4004599259 & 0.395840334 \\
  15 & -0.11933878407038 & 0.417644431 & 0.4003943056 & 0.396277293 \\
  16 & -0.11933880882938 & 0.417543502 & 0.4003411561 & 0.396645072 \\
  17 & -0.11933882639297 & 0.417458199 & 0.4002975557 & 0.396957882 \\
  18 & -0.11933883909801 & 0.417385385 & 0.4002613844 & 0.397226416 \\
  19 & -0.11933884844969 & 0.417322683 & 0.4002310753 & 0.397458855 \\
  \hline
  $\infty$ & -0.11933888188(1) & 0.416667(1) & 0.3999994(9) & 0.399999(1) \\
  \hline
\end{tabular}
\end{center}
\end{table}

%%%%%%%%%%%%%%%%%%%%%%%%%%%%%%%%%%%%%%%%%%%%%%%%%%%%%%%%%%%%%%%%%%%%
\section{Phenomenological Renormalization}

The largest and the second-largest eigenvalues of the transfer matrix
are calculated by the power method~\cite{Wilkinson65} up to $L=28$ and
$L=20$ for the row-to-row and diagonal-to-diagonal transfer matrices,
respectively.  In the present case, other improved variations of the
power method such as the conjugate-gradient method or the Lancz{\'o}s
method~\cite{StoerB92} are found to be less numerically stable.  In
the computation, we use quadratic-precision real numbers instead of
double-precision numbers, so that finite-size quantities ($z_L^*$ in
eq.~(\ref{eqn:rm}), etc.) can be obtained with the uncertainty of
$10^{-20}$ or less.  As seen below, finite-size data exhibit quite
rapid convergence to the thermodynamic limit.  Then, the quite high
accuracy for them is required to make meaningful extrapolations.

The critical point is estimated by
solving the phenomenological renormalization equations:
\begin{equation}
  \label{eqn:rm}
  L g_L (z^*_L) = L' g_{L'} (z^*_L)
\end{equation}
for two successive system sizes $L$ and $L'=L+1$.  As shown in Table~I
and II, there is no alternating behavior in finite-size corrections
with respect to the parity of $L$. It is in contrast with the case for
the physical critical point at $z=z_{\rm c}^+$, where the systems with
even and odd width have different finite-size corrections with each
other, reflecting the anti-ferromagnetic-like $\sqrt{2}\times\sqrt{2}$
ordering in the ordered phase~\cite{TodoS96}. Then, iterated
fits~\cite{BloteW90} of 3rd order yields
\begin{equation}
  \label{eqn:zcr}
  z_{\rm c}^- = -0.11933888189(2) \\
\end{equation}
for the row-to-row transfer matrix and
\begin{equation}
  z_{\rm c}^- = -0.11933888188(1) \\
\end{equation}
for the diagonal-to-diagonal one.  They agree excellently with each
other.  They are also consistent with eq.~(\ref{eqn:guttmann}), but
much accurate by about two orders of magnitude.

The exponent $\nu$ can be found from derivatives of the gap:
\begin{equation}
  1/\nu_L = \frac{\log \left( \left.
      \frac{d g_{L'}}{dz} \right/ \frac{d g_L}{dz}
  \right)}{\log(L'/L)} + 1
\end{equation}
at $z=z_L^*$.  The results are also listed in Table~I and II.  They
are extrapolated to the thermodynamic limit by iterated fits as
\begin{equation}
  \nu = 0.416668(2)
\end{equation}
and
\begin{equation}
  \nu = 0.416667(1)
\end{equation}
in the row-to-row and diagonal-to-diagonal cases, respectively.  These
results indicate $\nu=\frac{5}{12}$, which is expected to be exact for
the Yang-Lee edge singularity in two dimensions~\cite{Cardy85}.

At the critical point, the scaled gap $L g_L(z)$ in eq.~(\ref{eqn:rm})
is believed to be universal~\cite{PrivmanF84,Cardy84}, that is,
\begin{equation}
  L g_L(z_{\rm c}^-) = 2 \pi x_\sigma
\end{equation}
for $L\rightarrow\infty$, where $x_\sigma$ would be the scaling
dimension of the dominant scaling operator and relate to the exponent
of the correlation function $\eta$ by $\eta=2 x_\sigma$.  It is also
expected that the coefficient of the leading finite-size correction to
the critical free-energy density becomes
universal~\cite{BloteCN86,Affleck86}:
\begin{equation}
  f_L(z_{\rm c}^-) = f_0 - \frac{\pi c}{6L^2} + \cdots ,
\end{equation}
where $f_0$ is the critical free-energy density of the bulk, and $c$
is the central charge of the Virasoro algebra, which characterize the
critical theory of the model~\cite{BelavinPZ84}.

In practice, finite-size estimates for $x_\sigma$ and $c$ are
calculated by
\begin{equation}
  \label{eqn:xs}
  x_{\sigma,L} = \frac{1}{2\pi} L g_L (z_L^*)
\end{equation}
and
\begin{equation}
  \label{eqn:c}
  c_L = \frac{6}{\pi}
  \frac{L^2 L^{'2}}{(L^{'2}-L^2)} \{f_{L'} (z_L^*)-f_L (z_L^*)\} ,
\end{equation}
respectively.  As seen in Table~\ref{tab:row} and \ref{tab:diagonal},
both of $x_{\sigma,L}$ and $c_L$ tend to rapidly converge to the same
value $\frac{2}{5}$ for $L\rightarrow\infty$.  The best estimates are
\begin{equation}
  c=0.399999(1)
\end{equation}
and
\begin{equation}
  x_\sigma=0.3999996(7) ,
\end{equation}
respectively.  Following these results, one may conclude that the
singularity at $z=z_{\rm c}^-$ would be described by a conformal field
theory with $c=\frac{2}{5}$ and $x_\sigma=\frac{2}{5}$.  However, this
theory does not appear in the classification of the unitary minimal
series by Friedan, Qiu and Shenker~\cite{FriedanQS84} ($c=1-6/m(m+1)$
with $m=3, 4, 5,\cdots$).  Is it an absolutely new class of
universality?

Another interpretation for our numerical results are possible.
Suppose the largest eigenvalue $\lambda_1$ corresponds {\em not} to
the identity operator but to the primary operator with a {\em
  negative} dimension. Then, the second largest one $\lambda_2$ now
corresponds to the identity operator with the dimension 0.  Accepting
this assumption, one should calculate the {\em true} central charge
and scaling dimension (say $\tilde{c}$ and $\tilde{x}_\sigma$,
respectively) same as eqs.~(\ref{eqn:xs}) and (\ref{eqn:c}), {\em but}
using the modified free-energy density and gap~\cite{ItzyksonSZ86}:
\begin{equation}
  \label{eqn:free2}
  \tilde{f}_L = -\frac{1}{L} \log \lambda_2
\end{equation}
and
\begin{equation}
  \label{eqn:gap2}
  \tilde{g}_L = \log \left| \frac{\lambda_2}{\lambda_1} \right|
\end{equation}
instead of those defined in eqs.~(\ref{eqn:free}) and (\ref{eqn:gap}).
In the expressions the largest and second-largest eigenvalues,
$\lambda_1$ and $\lambda_2$ change places with each other in
comparison with the former definitions.  Thus, $\tilde{c}$ and
$\tilde{x}_\sigma$ are related with $c$ and $x_\sigma$ above mentioned
as
\begin{eqnarray}
  \tilde{c} &=& c - 12 x_\sigma \nonumber \\
  &=& -4.399996(8)
\end{eqnarray}
and
\begin{eqnarray}
  \label{eqn:res_x}
  \tilde{x}_\sigma &=& -x_\sigma \nonumber \\
  &=& -0.3999996(7) ,
\end{eqnarray}
respectively.  These values coincides quite excellently with Cardy's
conjecture~\cite{Cardy85} that the Yang-Lee edge singularity would be
described by the non-unitary conformal field theory with
$c=-\frac{22}{5}$ and $x_\sigma=-\frac{2}{5}$.  It is also confirmed
that $\nu=\frac{5}{12}$ and $\eta=2 \tilde{x}_\sigma=-\frac{4}{5}$
satisfy Fisher's scaling relation~\cite{Fisher78} $\nu=2/(d+2-\eta)$,
where $d$ is the dimensionality of space.  In the next section, we
investigate the critical eigenvalue spectrum of the transfer matrix in
order to confirm our tentative conclusion more precisely.

\begin{table}[t]
\caption{Eigenvalue spectrum of the transfer matrix at $z=z_{\rm c}^-$.
  The degeneracy of each level is denoted by $d$ in the second
  column.  The degeneracy and the scaling dimensions of the
  operators appearing in the modular invariant partition function
  for the Yang-Lee edge singularity~\cite{ItzyksonSZ86} are also
  listed in the fourth and fifth columns, respectively.}
\label{tab:spectrum}
\begin{center}
\begin{tabular}{lclll}
  \hline
  & \multicolumn{2}{l}{\ \ \ Present result} & 
  \multicolumn{2}{l}{\ \ \ Reference~\cite{ItzyksonSZ86}} \\
  \ $k$ & d & \ \ \ $\tilde{x}_k$ & d & \ \ \ $\tilde{x}_k$ \\
  \hline 
  1     & 1 & -0.3999996(7) & 1 & $-\frac{2}{5}$
  \vspace{0.5\tabcolsep} \\
  2     & 1 & \ 0           & 1 & $ \ \ \, 0 $
  \vspace{0.5\tabcolsep} \\
  3 4   & 2 & \ 0.6001(2)   & 2 & $-\frac{2}{5} + 1$
  \vspace{0.5\tabcolsep} \\
  5     & 1 & \ 1.599(2)    & 
  \raisebox{-0.6\tabcolsep}[0cm][0cm]{3} & 
  \raisebox{-0.6\tabcolsep}[0cm][0cm]{$-\frac{2}{5} + 2$} \\
  6 7   & 2 & \ 1.598(3)
  \vspace{0.5\tabcolsep} \\
  8 9   & 2 & \ 2.0001(1)   & 2 & $ \ \ \, 0 + 2$
  \vspace{0.5\tabcolsep} \\
  10 11 & 2 & \ 2.607(8)    &
  \raisebox{-0.6\tabcolsep}[0cm][0cm]{4} & 
  \raisebox{-0.6\tabcolsep}[0cm][0cm]{$-\frac{2}{5} + 3$} \\
  12 13 & 2 & \ 2.606(9)
  \vspace{0.5\tabcolsep} \\
  14 15 & 2 & \ 2.996(8)    & 2 & $ \ \ \, 0 + 3$ \\
  \hline
\end{tabular}
\end{center}
\end{table}

%%%%%%%%%%%%%%%%%%%%%%%%%%%%%%%%%%%%%%%%%%%%%%%%%%%%%%%%%%%%%%%%%%%%
\section{Eigenvalue Spectrum of Transfer Matrix at Critical Point}

The complete spectrum of the transfer matrix at the critical point
contains more information about the criticality.  It is known that at
the critical point, there is asymptotically one-to-one correspondence
between the eigenvalues of the transfer matrix and the operators
appearing in the corresponding conformal field
theory~\cite{Cardy87,Cardy84}.

We calculated all the eigenvalues of the row-to-row transfer matrix
(\ref{eqn:tm_rr}) at $z=z_{\rm c}^-$ by the combination of the
Householder transformation and the QR algorithm~\cite{Wilkinson65}.
The corresponding scaling dimensions are then calculated as
\begin{equation}
  \tilde{x}_k = \lim_{L\rightarrow\infty} 
  \frac{1}{2\pi} L \tilde{g}_L^{(k)} (z_{\rm c}^-)
\end{equation}
with
\begin{equation}
  \label{eqn:gap3}
  \tilde{g}_L^{(k)} = \log \left| \frac{\lambda_2}{\lambda_k} \right|,
\end{equation}
where $\lambda_k$ is the $k$th-largest eigenvalue.  Note that the gap
in eq.~(\ref{eqn:gap3}) is measured from the {\em second}-largest
eigenvalue $\lambda_2$ following the assumption in the last section.

The extrapolated values for the first 15 eigenvalues
($k=1,2,\cdots,15$) are listed in Table III with their degeneracy.
(The results for $\tilde{x}_1$ has been obtained as $\tilde{x}_\sigma$
in the last section (eq.~(\ref{eqn:res_x})), and $\tilde{x}_2$ is
apparently 0 by definition (eq.~(\ref{eqn:gap3})).)

First, it should be noted that the dimension 1 is absent in the
eigenvalue spectrum (Table~III).  This supports our conclusion,
because the level-1 descendants of the identity operator generally
vanish~\cite{BelavinPZ84}. On the other hand, the dimension
$-\frac{2}{5}+1$ is present, which corresponds to the level-1
secondary operators of the operator with the dimension
$\tilde{x}_1=\tilde{x}_\sigma$.

Moreover, the whole eigenvalue spectrum, at least up to $k=15$,
coincides completely with the operator content of the modular
invariant partition function for $c=-\frac{22}{5}$ conformal theory
obtained by Itzykson {\it et al.}~\cite{ItzyksonSZ86}, including the
degeneracy. (Their results are also listed in the fourth and fifth
columns of Table~III).

%%%%%%%%%%%%%%%%%%%%%%%%%%%%%%%%%%%%%%%%%%%%%%%%%%%%%%%%%%%%%%%%%%%%
\section{Summary and Discussion}

In the present paper, we investigated the hard-core singularity of the
hard-square lattice gas by the transfer-matrix method and the
phenomenological renormalization.  In special, for the row-to-row
transfer matrix we perform diagonalization up to the quite large
system width $L=28$ by using the improved sparse-matrix factorization
technique.  Finite-size analysis for the two different transfer
matrices yields the highly consistent results with each other.  In
terms of the phenomenological renormalization technique, we obtained
the accurate estimate for the critical point $z_{\rm c}$, which is
consistent with the previous studies but much accurate.

The exponent $\nu$, the central charge $c$ and the dominant scaling
dimension $x_\sigma$ are also determined accurately.  The results
indicate strongly that the hard-core singularity is described by the
{\em non-unitary} conformal field theory with $c=-\frac{22}{5}$, in
other words, it belongs to the same universality class as the Yang-Lee
edge singularity.

As far as we know, the present result for the critical point is one of
the most accurate numerical estimates by means of the computer
simulation for the non-trivial problems in statistical physics.  The
rapidness of convergence or the smallness of finite-size corrections
would be due to the extreme simplicity of the critical theory ---
there are only two primary operators ($x=0$ and $-\frac{2}{5}$).

Finally, we comment that the hard-square lattice gas can be considered
as a special limit of the Ising anti-ferromagnet on the square
lattice~\cite{LeeY52} ($\beta J \rightarrow - \infty$ and $\beta H
\rightarrow \infty$ with $\beta(H+2J)$ fixed, where $\beta$ is the
inverse temperature, $J$ the nearest-neighbor anti-ferromagnetic
interaction ($J<0$) and the $H$ is the magnetic field). In the latter
model, the Yang-Lee edges are expected to form a couple of critical
{\em lines} in the complex $H$ plane above the critical temperature.
The connection between them is an interesting problem, but it remains
as an open question.

%%%%%%%%%%%%%%%%%%%%%%%%%%%%%%%%%%%%%%%%%%%%%%%%%%%%%%%%%%%%%%%%%%%%
\section*{Acknowledgment}

The present author would like to thank Prof.~H. Takayama for critical
reading of the manuscript.  The computation in this work has been
performed using the facilities of the Supercomputer Center, Institute
for Solid State Physics, University of Tokyo.

\end{document}